\documentclass[useAMS]{mn2e}
 \usepackage{times}
 \usepackage{graphicx}
 \usepackage{epsfig}
 \usepackage{amssymb}

 \title[A transient Trojan companion to Venus]
       {Asteroid 2013 ND$_{15}$: Trojan companion to Venus, PHA to the Earth}

 \author[C. de la Fuente Marcos and R. de la Fuente Marcos]
        {C.~de~la~Fuente~Marcos\thanks{E-mail: nbplanet@fis.ucm.es}
          and
         R. de la Fuente Marcos \\
         Universidad Complutense de Madrid,
         Ciudad Universitaria, E-28040 Madrid, Spain}
 \date{Accepted 2014 January 20.
       Received 2014 January 19;
       in original form 2013 October 18}
 \pubyear{2014}
 
\begin{document}
  \maketitle

  \begin{abstract}
     Venus has three known co-orbitals: (322756)~2001~CK$_{32}$, 
     2002~VE$_{68}$ and 2012~XE$_{133}$. The first two have absolute 
     magnitudes 18~$<H<$~21. The third one, significantly smaller at
     $H$~=~23.4 mag, is a recent discovery that signals the probable 
     presence of many other similar objects: small transient 
     companions to Venus that are also potentially hazardous asteroids 
     (PHAs). Here, we study the dynamical evolution of the recently 
     discovered asteroid 2013~ND$_{15}$. At $H$~=~24.1 mag, this minor 
     body is yet another small Venus co-orbital and PHA, currently 
     close to the Lagrangian point L$_4$ and following the most 
     eccentric path found so far for objects in this group. This 
     transient Trojan will leave the 1:1 mean motion resonance within 
     a few hundred years although it could be a recurrent librator. 
     Due to its high eccentricity (0.6), its dynamics is different 
     from that of the other three known Venus co-orbitals even if they 
     all are near-Earth objects (NEOs). A Monte Carlo simulation that 
     uses the orbital data and discovery circumstances of the four 
     objects as proxies to estimate the current size of this 
     population, indicates that the number of high-eccentricity, 
     low-inclination Venus co-orbital NEOs may have been greatly 
     underestimated by current models. Three out of four known objects 
     were discovered with solar elongation at perigee greater than 
     135\degr even if visibility estimates show that less than 4 per 
     cent of these objects are expected to reach perigee at such large 
     elongations. Our calculations suggest that the number of minor 
     bodies with sizes above 150 m currently engaged in co-orbital 
     motion with Venus could be at least one order of magnitude larger 
     than usually thought; the number of smaller bodies could easily 
     be in many thousands. These figures have strong implications on 
     the fraction of existing PHAs that can barely be detected by 
     current surveys. Nearly 70 per cent of the objects discussed here 
     have elongation at perigee $<$~90\degr and 65 per cent are 
     prospective PHAs. 
  \end{abstract}

  \begin{keywords}
     celestial mechanics --
     minor planets, asteroids: individual: 2002 VE$_{68}$ -- 
     minor planets, asteroids: individual: 2012 XE$_{133}$ --
     minor planets, asteroids: individual: 2013 ND$_{15}$ --
     minor planets, asteroids: individual: 322756 -- 
     planets and satellites: individual: Venus. 
  \end{keywords}

  \section{Introduction}
     Venus has no known satellites larger than about 0.3 km in radius (Sheppard \& Trujillo 2009). As for unbound companions, multiple 
     secular resonances make the presence of primordial (long-term stable) Venus co-orbitals very unlikely (Scholl, Marzari \& Tricarico 
     2005). In contrast, numerical simulations consistently show that present-day temporary co-orbital companions to Venus should exist 
     (Mikkola \& Innanen 1992; Michel 1997; Christou 2000; Tabachnik \& Evans 2000; Brasser \& Lehto 2002). This theoretical prospect has 
     been confirmed with the identification of (322756) 2001 CK$_{32}$, a horseshoe-quasi-satellite orbiter (Brasser et al. 2004), 2002 
     VE$_{68}$, a quasi-satellite (Mikkola et al. 2004) and 2012~XE$_{133}$, a former L$_5$ Trojan moving into a horseshoe orbit (de la 
     Fuente Marcos \& de la Fuente Marcos 2013). Besides being temporarily trapped in a 1:1 mean motion resonance with Venus, all these 
     objects are Mercury grazers, Venus crossers and Earth crossers. As Earth crossers, they are also members of the near-Earth object (NEO) 
     population that experiences close approaches to our planet. The minimum orbit intersection distances (MOIDs) with the Earth of 322756, 
     2002 VE$_{68}$ and 2012~XE$_{133}$ are 0.08, 0.03 and 0.003 au, respectively. This property makes them objects of significant interest. 
     \hfil\par
     Potentially hazardous asteroids (PHAs) are currently defined as having an Earth MOID of 0.05 au or less and an absolute magnitude, $H$, 
     of 22.0 or less (Marsden 1997); following this definition and at the time of this writing, there are 1451 known 
     PHAs.\footnote{http://neo.jpl.nasa.gov/neo/groups.html} If the size limit ($>$ 150 m, for an assumed albedo of 13 per cent) is lowered 
     or ignored (hereafter, the PHA regime), there are several thousands of known PHAs. Lowering the size limit appears to be well supported 
     by the analysis of the effects of the recent impact of a small asteroid in central Russia, the Chelyabinsk Event (see e.g. Brown et al. 
     2013), which demonstrates that even smaller objects (under 20~m) can still cause significant damage if they hit densely populated 
     areas. In any case, asteroids 322756 and 2002 VE$_{68}$ have absolute magnitudes 18~$<H<$~21; therefore, 2002 VE$_{68}$ is a PHA in 
     strict sense according to the current definition. 
     \hfil\par
     Discarding a primordial origin for dynamical reasons, the genesis of this group of NEOs is far from well understood. These objects 
     exhibit resonant (or near-resonant) behaviour with Mercury (9:23), Venus (1:1) and the Earth (8:13), following rather chaotic paths 
     with $e$-folding times of the order of 100 yr. Close encounters with the three innermost planets make this population intrinsically 
     transient; therefore, one or more mechanisms should be at work to mitigate the losses, repopulating the group. Christou (2000) 
     predicted a quasi-steady-state flux of minor bodies moving in and out of the co-orbital state with Venus. Morais \& Morbidelli (2006) 
     explained the existence of this population within the steady-state scenario envisioned by Bottke et al. (2000, 2002) where NEOs are 
     constantly being supplied from the main asteroid belt. Predictions from their model appear to be consistent with the available 
     observational evidence but they imply that current surveys have already reached completeness for $H <$ 21. However, some crucial 
     observational facts have not yet been taken into consideration. For instance, how efficient are ongoing surveys expected to be in the 
     case of objects that spend most of the time at solar elongations $<$ 90\degr? Are the known objects just the tip of the iceberg, 
     representing an unexpected abundance of small minor bodies whose orbits make them companions to Venus and PHAs to the Earth?  
     \hfil\par
     In this paper, we show that the recently discovered asteroid 2013 ND$_{15}$ is another small transient Venus co-orbital in the PHA 
     regime (with a MOID of 0.007 au, but an $H$ = 24.1 mag). Together with 2012 XE$_{133}$ ($H$= 23.4 mag), they may signal the presence of 
     a significant population of small transient Venus co-orbitals in the PHA regime. To explore this interesting possibility, the orbital 
     properties of these objects and their discovery circumstances can be used to estimate (i) their chances of being found by ongoing NEO 
     surveys and, indirectly, (ii) the actual size of this population. This paper is organized as follows. In Section 2, we present the 
     available data on 2013 ND$_{15}$ and briefly outline our numerical model. Section 3 focuses on analysing the past, present and future 
     dynamical evolution of the object. Section 4 provides a comparative dynamical analysis with the other three Venus co-orbitals. Some 
     relevant issues regarding the observability of these objects are discussed in Section 5. Our conclusions are summarized in Section 6.
%
%
     \begin{table*}
      \fontsize{8}{11pt}\selectfont
      \tabcolsep 0.2truecm
      \caption{Heliocentric Keplerian orbital elements (and the 1$\sigma$ uncertainty) of asteroids 2013 ND$_{15}$, (322756) 2001 CK$_{32}$, 
               2002 VE$_{68}$ and 2012 XE$_{133}$ (Epoch = JD2456600.5, 2013-Nov-4.0; J2000.0 ecliptic and equinox. Source: JPL Small-Body 
               Database).
              }
      \begin{tabular}{cccccc}
       \hline
                                                               &   &   2013 ND$_{15}$      &   322756                      &   
                                                                       2002 VE$_{68}$                &   2012 XE$_{133}$     \\
       \hline
        Semimajor axis, $a$ (au)                               & = &   0.7235$\pm$0.0002   &   0.725457880$\pm$0.000000004 &   
                                                                       0.7236543902$\pm$0.0000000005 &   0.72297$\pm$0.00014 \\
        Eccentricity, $e$                                      & = &   0.6115$\pm$0.0006   &   0.3824353$\pm$0.0000002     &   
                                                                       0.41033045$\pm$0.00000005     &   0.4332$\pm$0.0003   \\
        Inclination, $i$ ($^{\circ}$)                          & = &   4.794$\pm$0.009     &   8.13201$\pm$0.00002         &   
                                                                       9.005960$\pm$0.000013         &   6.711$\pm$0.008     \\
        Longitude of the ascending node, $\Omega$ ($^{\circ}$) & = &  95.84$\pm$0.02       & 109.5014$\pm$0.0002           & 
                                                                     231.579688$\pm$0.000005         & 281.095$\pm$0.006     \\
        Argument of perihelion, $\omega$ ($^{\circ}$)          & = &  19.69$\pm$0.02       & 234.0956$\pm$0.0002           & 
                                                                     355.463807$\pm$0.000014         & 337.085$\pm$0.007     \\
        Mean anomaly, $M$ ($^{\circ}$)                         & = & 357.57$\pm$0.06       &   1.35441$\pm$0.00009         & 
                                                                     123.30681$\pm$0.00003           & 351.6$\pm$0.2         \\
        Perihelion, $q$ (au)                                   & = &   0.2811$\pm$0.0005   &   0.44801718$\pm$0.00000011   &   
                                                                       0.42671696$\pm$0.00000004     &   0.4098$\pm$0.0003   \\
        Aphelion, $Q$ (au)                                     & = &   1.1660$\pm$0.0002   &   1.002898585$\pm$0.000000006 &   
                                                                       1.0205591818$\pm$0.0000000007 &   1.0361$\pm$0.0002   \\
        Absolute magnitude, $H$ (mag)                          & = &  24.1                 &  18.9                         &  
                                                                      20.5                 &  23.4                           \\
       \hline
      \end{tabular}
      \label{elements}
     \end{table*}
%
%

  \section{Data and numerical model}
     Asteroid 2013 ND$_{15}$ was discovered by N. Primak, A. Schultz, T. Goggia and K. Chambers observing for the Pan-STARRS 1 project from 
     Haleakala on 2013 July 13 (Wainscoat et al. 2013). The object was first observed using a 1.8 m Ritchey--Chretien telescope at an 
     apparent $w$ magnitude of 20.5; it was re-observed in subsequent nights from Cerro Tololo Observatory and with the 3.6 m 
     Canada--France--Hawaii Telescope from Mauna Kea. With a value of the semimajor axis $a$ = 0.7235 au, very close to that of Venus 
     (0.7233 au), this Aten asteroid is an NEO moving in a very eccentric, $e$ = 0.61, and a little inclined, $i = 4\fdg8$, orbit that makes 
     it a Mercury, Venus and Earth crosser. Therefore, its orbit is slightly different from those of the three previously known Venus 
     co-orbitals (see Table \ref{elements}): (322756) 2001 CK$_{32}$, 2002 VE$_{68}$ and 2012 XE$_{133}$. The source of the Heliocentric 
     Keplerian osculating orbital elements and uncertainties in Table \ref{elements} is the Jet Propulsion Laboratory (JPL) Small-Body 
     Database.\footnote{http://ssd.jpl.nasa.gov/sbdb.cgi} 
     \hfil\par
     Its very small relative semimajor axis, $|a - a_{\rm Venus}| \sim$ 0.0002$\pm$0.0002 au (the smallest found so far), makes this object 
     a clear candidate to be a Venus co-orbital. It completes one orbit around the Sun in 224.79 d or 0.6154 yr when Venus does it in 
     224.70 d or 0.6152 yr. Its current orbit is based on 21 observations with a data-arc span of 26 d. As expected of recent discoveries, 
     the quality of the orbits of both 2012 XE$_{133}$ and 2013 ND$_{15}$ is currently lower than that of the other two co-orbitals. It is 
     similar, however, to the one available when the other two objects were identified as unbound companions to Venus back in 2004 (de la 
     Fuente Marcos \& de la Fuente Marcos 2013). This object has $H$ = 24.1 mag (assumed $G$ = 0.15) or a diameter of 40 to 100 m for an 
     assumed albedo in the range 0.20--0.04.
     \hfil\par
     As a Venus co-orbital candidate, the key parameter to study is the oscillation of the difference between the mean longitudes of the 
     object and Venus or relative mean longitude, $\lambda_{\rm r}$. The mean longitude of an object is given by $\lambda$ = $M$ + $\Omega$ 
     + $\omega$, where $M$ is the mean anomaly, $\Omega$ is the longitude of the ascending node and $\omega$ is the argument of perihelion. 
     An object is co-orbital if $\lambda_{\rm r}$ oscillates (librates) around a constant value; if $\lambda_{\rm r}$ can take any value 
     (circulates) then we have a passing object. For additional details on co-orbital dynamical behaviour see e.g. Namouni (1999) and 
     Christou (2000). 
     \hfil\par
     To confirm the co-orbital nature of 2013 ND$_{15}$ with Venus, we have carried out $N$-body simulations using the Hermite integrator 
     (Makino 1991; Aarseth 2003) in both directions of time for 10 kyr. The standard version of this direct $N$-body code is publicly 
     available from the IoA web site.\footnote{http://www.ast.cam.ac.uk/$\sim$sverre/web/pages/nbody.htm} Relativistic terms and the role of 
     the Yarkovsky and Yarkovsky--O'Keefe--Radzievskii--Paddack (YORP) effects (see e.g. Bottke et al. 2006) are neglected. The non-inclusion 
     of these effects has no impact on the assessment of the current dynamical status of this object but may affect predictions regarding 
     its future evolution and dynamical past. In particular, the Yarkovsky effect may have a non-negligible role on the medium- and 
     long-term evolution of objects as small as 2013 ND$_{15}$. Proper modelling of the Yarkovsky force requires knowledge on the physical 
     properties of the objects involved (for example, rotation rate, albedo, bulk density, surface conductivity, emissivity) which is not 
     the case for 2013 ND$_{15}$. Detailed observations obtained during future encounters with the Earth or from the {\it Gaia} mission 
     should be able to provide that information and eventually improve the modelling presented here (see e.g. Carbognani et al. 2012; 
     Cellino \& Dell'Oro 2012; Delb\'o et al. 2012).
     \hfil\par
     Our model Solar system includes the perturbations by the eight major planets, treating the Earth and the Moon as two separate objects, 
     the barycentre of the dwarf planet Pluto--Charon system and the three largest asteroids. We use initial conditions (positions and 
     velocities in the barycentre of the Solar system referred to the JD2456600.5 epoch, $t$ = 0 coincides with this epoch) provided by the 
     JPL's \textsc{horizons} ephemeris system (Giorgini et al. 1996; Standish 1998).\footnote{http://ssd.jpl.nasa.gov/horizons.cgi} Relative 
     errors in the total energy at the end of the simulations are $< 2 \times 10^{-14}$. Additional details can be found in de la Fuente 
     Marcos \& de la Fuente Marcos (2012, 2013). 
     \hfil\par
     In addition to the calculations performed using the nominal orbital elements in Table \ref{elements}, we have completed 100 control 
     simulations using sets of orbital elements obtained from the nominal ones within the accepted uncertainties. The computed set of 
     control orbits follows a normal distribution in the six-dimensional orbital parameter space. These synthetic orbital elements are 
     compatible with the nominal ones within the 3$\sigma$ deviations (see Table \ref{elements}) and reflect the observational uncertainties 
     in astrometry. The short-term dynamical evolution of the entire set of control orbits is consistent with the one derived from the 
     nominal orbit. 
%
%
     \begin{figure}
       \centering
        \includegraphics[width=\linewidth]{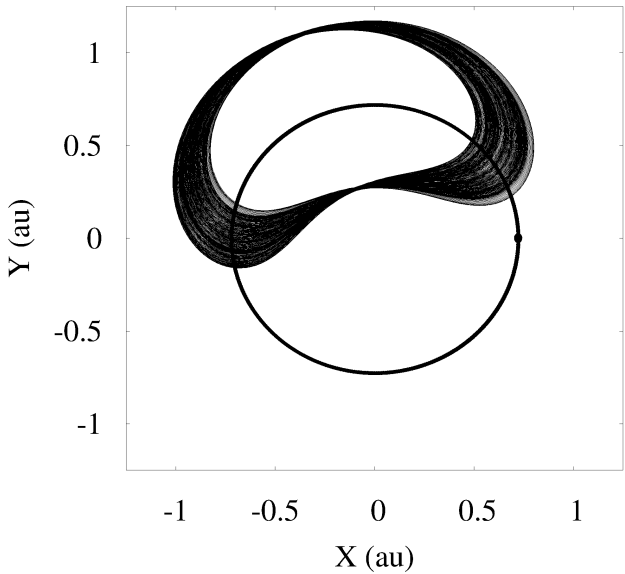}
        \caption{The motion of 2013 ND$_{15}$ over the time range (-200, 50) yr is displayed projected on to the ecliptic plane in a 
                 coordinate system rotating with Venus (nominal orbit in Table \ref{elements}). The orbit and position of Venus are also 
                 indicated. 
                }
        \label{trojan}
     \end{figure}
%
%

  \section{Current dynamical status and evolution}
\begin{figure*}
       \centering
        \includegraphics[width=\linewidth]{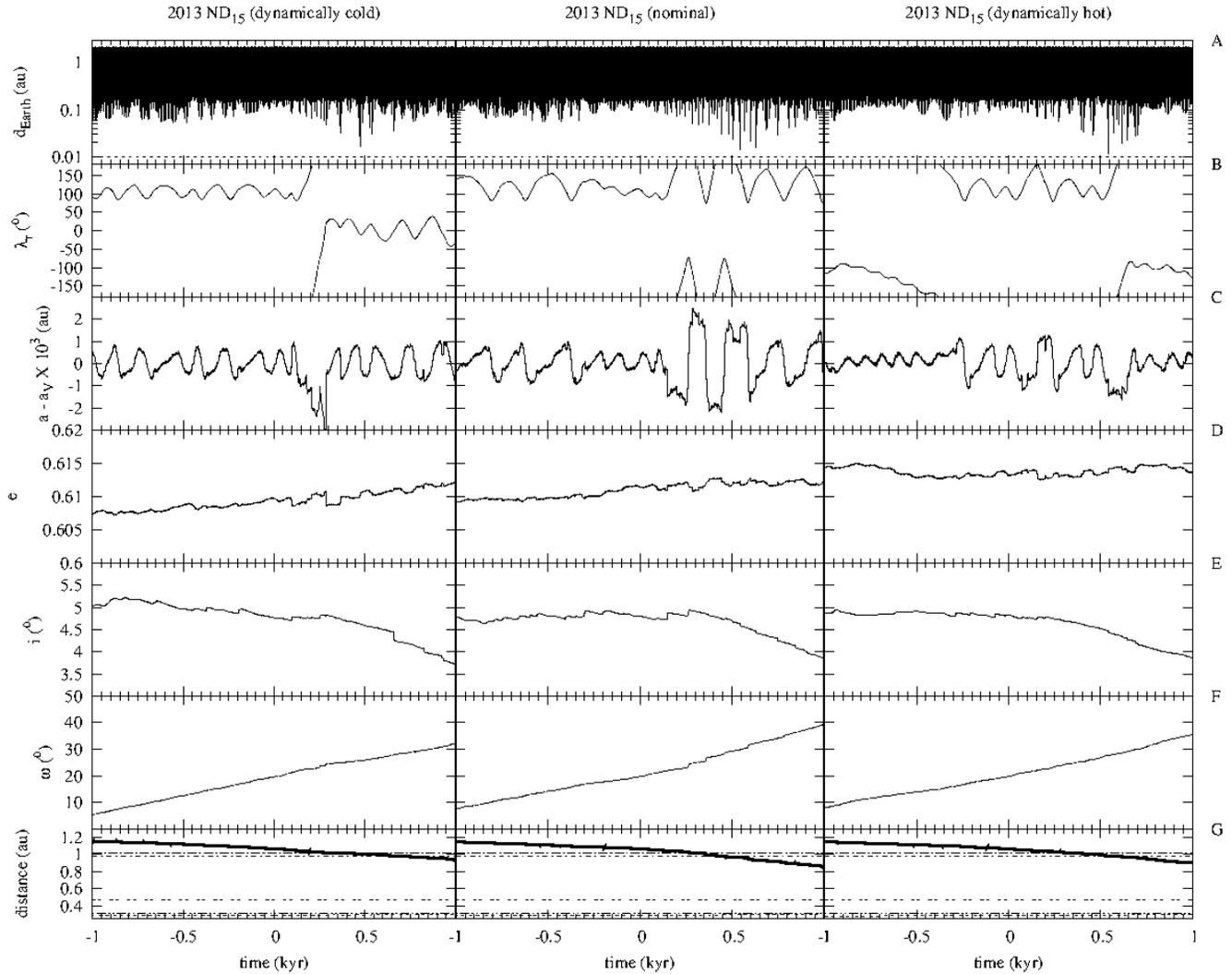}
        \caption{Comparative short-term dynamical evolution of various parameters for an orbit arbitrarily close to the nominal one of 
                 2013~ND$_{15}$ as in Table \ref{elements} (central panels) and two representative examples of orbits that are most 
                 different from the nominal one (see the text for details). The distance from the Earth (A-panels); the value of the Hill 
                 sphere radius of the Earth, 0.0098 au, is displayed. The resonant angle, $\lambda_{\rm r}$ (B-panels). The orbital elements 
                 $a - a_{\rm Venus}$ (C-panels), $e$ (D-panels), $i$ (E-panels) and $\omega$ (F-panels). The distances to the descending 
                 (thick line) and ascending nodes (dotted line) appear in the G-panels. Earth's and Mercury's aphelion and perihelion 
                 distances are also shown. 
                }
        \label{control}
     \end{figure*}

     The motion of 2013 ND$_{15}$ over the time range (-200, 50) yr as seen in a coordinate system rotating with Venus projected on to the 
     ecliptic plane is plotted in Fig. \ref{trojan} (nominal orbit in Table \ref{elements}). This minor body is a Venus co-orbital currently 
     following a tadpole orbit around Venus' Lagrangian point L$_4$; it is, therefore, a Trojan (see e.g. Murray \& Dermott 1999). Due to 
     its significant eccentricity and in accordance to theoretical predictions (Namouni, Christou \& Murray 1999; Namouni \& Murray 2000), 
     the libration angle is greater than the usual value of +60\degr. The libration centre is displaced from the typical equilateral 
     location for eccentric orbits; the short-term evolution of the resonant angle, $\lambda_{\rm r}$, is displayed in the B-panels of Fig. 
     \ref{control}. 
     \hfil\par
%
%
%
%
     All the integrated control orbits for 2013 ND$_{15}$ exhibit Trojan libration with respect to Venus at $t$ = 0. As an example, Fig. 
     \ref{control} displays the short-term dynamical evolution of an orbit arbitrarily close to the nominal one (central panels) and those 
     of two representative worst orbits which are most different from the nominal one. The orbit labelled as `dynamically cold' (left-hand 
     panels) has been obtained by subtracting thrice the uncertainty from the orbital parameters (the six elements) in Table \ref{elements}. 
     It is indeed the coldest, dynamically speaking, orbit (lowest values of $a$, $e$ and $i$) compatible with the current values of its 
     orbital parameters. In contrast, the orbit labelled as `dynamically hot' (right-hand panels) was computed by adding three times the 
     value of the uncertainty to the orbital elements in Table \ref{elements}. This makes this trajectory the hottest possible in dynamical 
     terms (largest values of $a$, $e$ and $i$). All the control orbits exhibit consistent behaviour within a few hundred years of $t = 0$. 
     Asteroid 2013 ND$_{15}$ is indeed a Venus Trojan according to the current observational uncertainties or at a confidence level $>$ 99 
     per cent. However, its Trojan dynamical status is only temporary and the tadpole orbit that it currently follows (see Fig. \ref{trojan}), 
     short lived.
     \hfil\par
%
%
     \begin{figure}
       \centering
        \includegraphics[width=\linewidth]{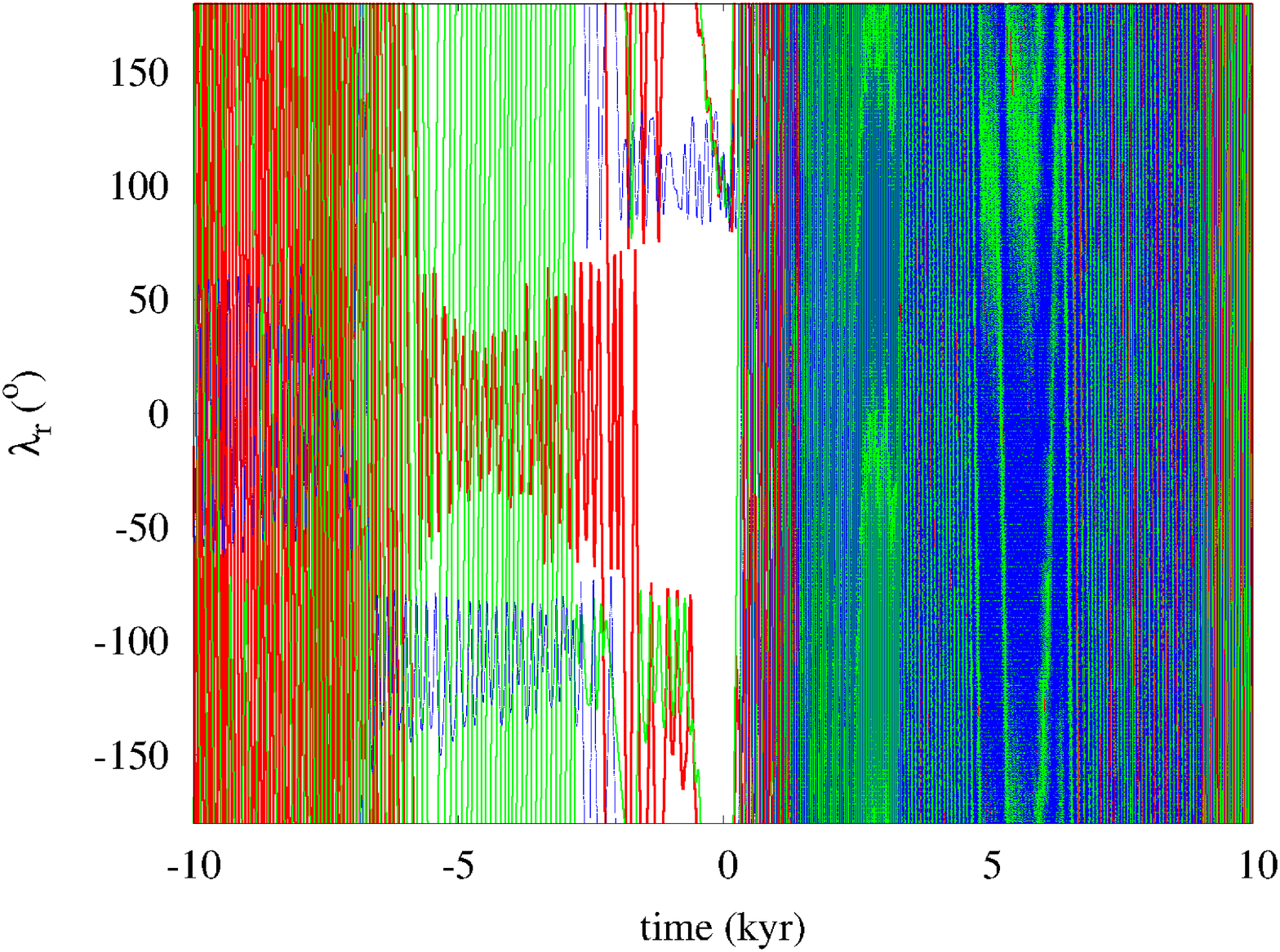}
        \caption{The resonant angle, $\lambda_{\rm r}$, for the nominal orbit of 2013 ND$_{15}$ in Table \ref{elements} (red line) and two 
                 of the control orbits. The blue line corresponds to a particular control orbit that was chosen close to the 3$\sigma$ limit 
                 so its orbital elements are most different from the nominal ones. The third orbit (green line) has osculating elements 
                 within 1$\sigma$ of those of the nominal orbit.  
                }
        \label{lambda}
     \end{figure}
%
%
     Our calculations (see Figs \ref{lambda} and \ref{all}, left-hand panels) show that 2013 ND$_{15}$ has already remained in the 1:1 
     commensurability with Venus for a few thousand years. During this time-span, the object may have been a Trojan (L$_4$ and L$_5$), 
     a quasi-satellite ($\lambda_{\rm r}$ = 0\degr) and a horseshoe librator ($\lambda_{\rm r}$ = 180\degr). Its future orbital evolution 
     strongly depends on the outcome of flybys with our planet during the next 250--500 yr. A very close encounter with the Earth at nearly 
     0.003 au may take place about 250 yr from now. Asteroid 2013 ND$_{15}$ may become a passing object after such an encounter but 
     recurrent co-orbital episodes are observed for most control orbits. This is consistent with its past orbital evolution. A future better 
     orbit may help to reduce the uncertainties in the predicted dynamical evolution but the fact is that much better orbits still translate 
     into rather chaotic orbital evolution both in the past and the future for (322756) 2001 CK$_{32}$ and 2002 VE$_{68}$ (see e.g. de la 
     Fuente Marcos \& de la Fuente Marcos 2012, 2013). For objects following intrinsically irregular paths, an improved orbit does not make 
     medium- or long-term predictions significantly better or more reliable.  

  \section{Asteroid 2013 ND$_{15}$ in context}
\begin{figure*}
       \centering
        \includegraphics[width=\linewidth]{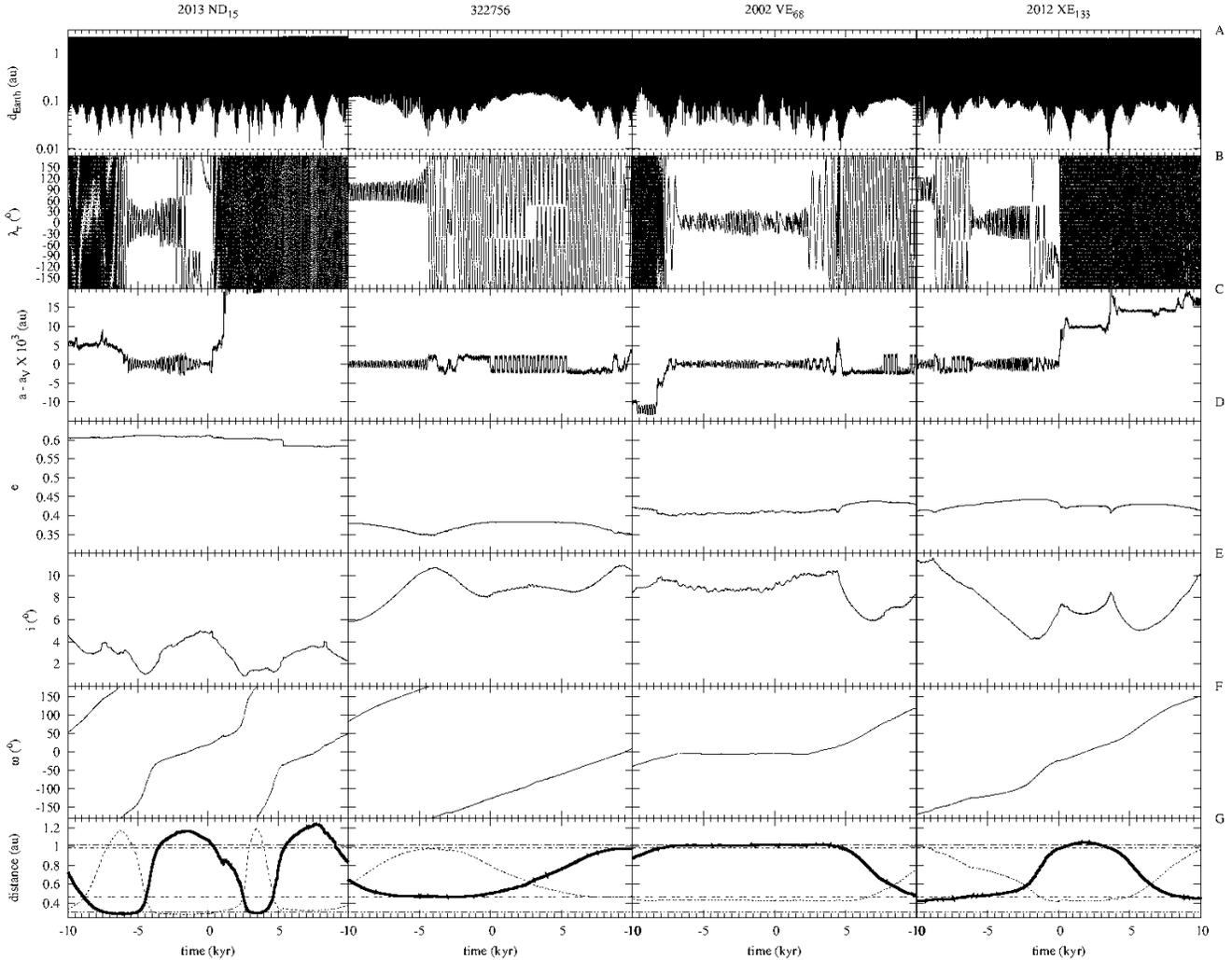}
        \caption{Comparative dynamical evolution of various parameters for the four known Venus co-orbitals. The distance from the Earth 
                 (panel A); the value of the Hill sphere radius of the Earth, 0.0098 au, is displayed. The resonant angle, $\lambda_{\rm r}$ 
                 (panel B) for the nominal orbit in Table \ref{elements}. The orbital elements $a - a_{\rm Venus}$ (panel C), $e$ (panel D), 
                 $i$ (panel E) and $\omega$ (panel F). The distances to the descending (thick line) and ascending nodes (dotted line) appear 
                 in panel G. Earth's and Mercury's aphelion and perihelion distances are also shown. 
                }
        \label{all}
     \end{figure*}

     Fig. \ref{all} displays the comparative evolution of the osculating orbital elements and other parameters of interest of all the 
     known Venus co-orbitals (nominal orbits in Table \ref{elements}). It is clear that the higher eccentricity (mainly) and slightly lower 
     inclination of the orbit of 2013 ND$_{15}$ are responsible for the obvious differences observed in the figure. One common feature is
     also evident, all these objects switch between resonant states relatively frequently. Transfers between tadpole, horseshoe and 
     quasi-satellite orbits are triggered by close encounters with the inner planets and those are the result of the libration of the nodes 
     (Wiegert, Innanen \& Mikkola 1998). Chaos-induced transfer from one Lagrange point to another appears to be a dynamical property
     common to temporarily captured Trojans of the terrestrial planets (Schwarz \& Dvorak 2012). 
     \hfil\par
     For an object following an inclined path, close encounters with major planets are only possible in the vicinity of the nodes. The 
     distance between the Sun and the nodes is given by $r = a (1 - e^2) / (1 \pm e \cos \omega)$, where the `+' sign is for the ascending 
     node and the `-' sign is for the descending node. The positions of the nodes are plotted in the G-panels. The evolution over time of 
     the location of the nodes of (322756) 2001 CK$_{32}$, 2002 VE$_{68}$ and 2012 XE$_{133}$ is very similar; they remain confined between 
     Mercury's and Earth's aphelia, 0.4667 au and 1.0167 au, respectively. In sharp contrast, the nodes of 2013 ND$_{15}$ exhibit a wider 
     oscillation range, 0.3--1.2 au. The lower inclination translates into more frequent flybys with the Earth. These encounters are not as 
     deep as in the case of some of the other three objects but they occur more often. Asteroid 2013 ND$_{15}$ seems to be an extreme 
     version of 2012 XE$_{133}$ and its orbit is even more unstable.

  \section{But, how many are out there?}
     The model in Morais \& Morbidelli (2006) only strictly applies if $H < 22$ and it predicts the existence of two Venus co-orbitals with 
     absolute magnitude brighter than 21 mag. There are two known objects within that limit: (322756) 2001 CK$_{32}$ and 2002 VE$_{68}$. 
     Morais \& Morbidelli (2006) already considered surprising that the currently known sample of these objects was apparently complete up 
     to $H < 21$. This result is even more extraordinary when we realize that ongoing NEO search programmes are not specifically targeting 
     this class of objects. A look back into the discovery circumstances of individual objects, as compiled in Table \ref{discovery}, may 
     help us to understand better this intriguing issue. In particular, the value of the solar elongation or angle between the Sun and the 
     object as seen from the Earth at the time of discovery can provide valuable clues on how abundant these objects really are. The source
     of the data in Table \ref{discovery} is the Minor Planet Center (MPC) Database.\footnote{http://www.minorplanetcenter.net/db\_search} 
     \hfil\par
%
%
\begin{figure}
       \centering
        \includegraphics[width=\linewidth]{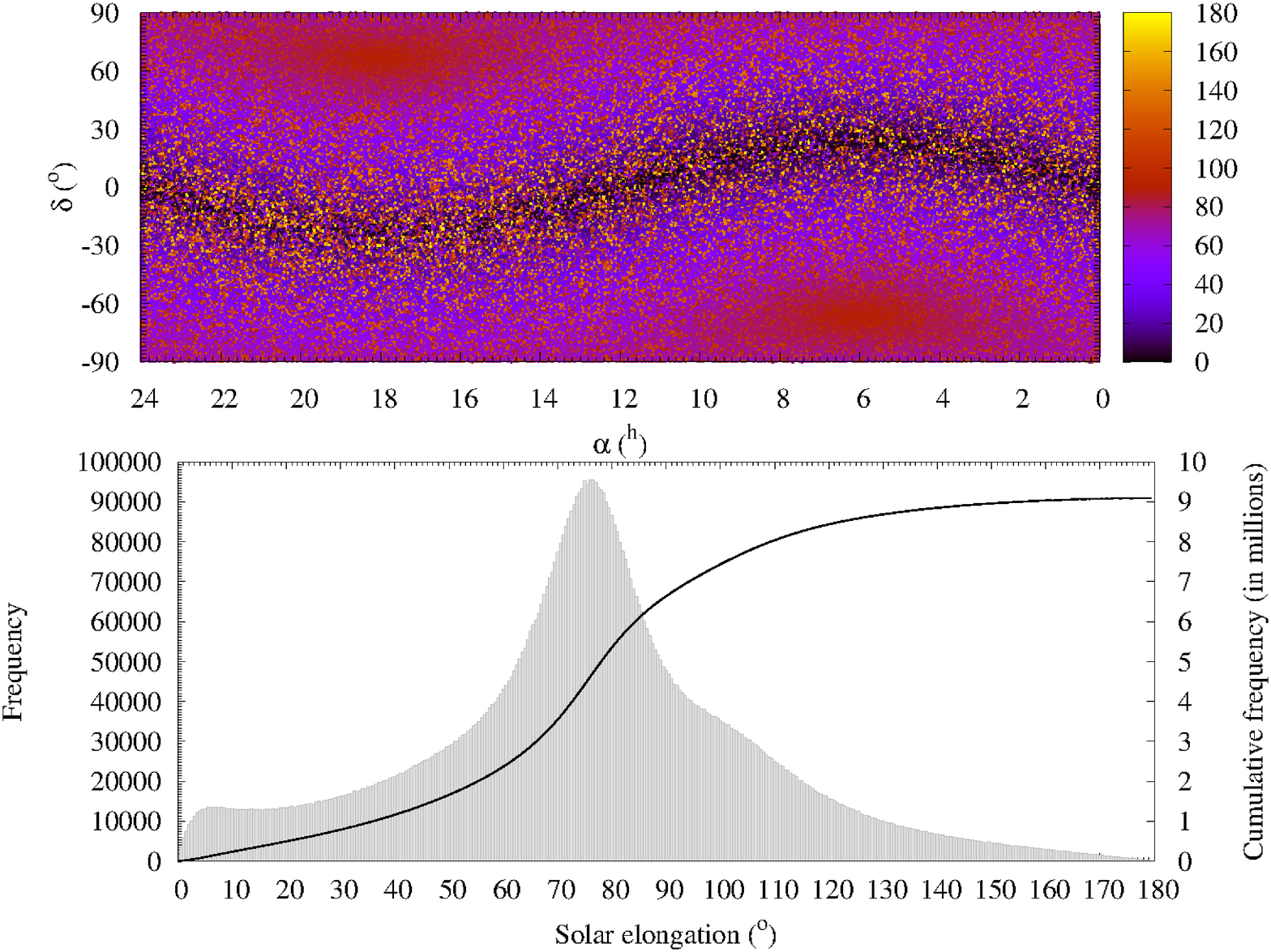}
        \caption{Distribution in equatorial coordinates at perigee of orbits similar to those of known Venus co-orbitals as a function of 
                 the (colour coded) geocentric solar elongation (top panel). A frequency analysis of the same data (bottom panel). Only 
                 perigees $<$ 0.1 au are considered.}
        \label{elongation}
     \end{figure}

%
%
%
%
         \begin{table}
          \centering
          \fontsize{8}{11pt}\selectfont
          \tabcolsep 0.12truecm
          \caption{Equatorial coordinates (J2000.0), apparent magnitudes (with filter), solar elongation, $\theta$, and phase, $\phi$, at 
                   discovery time of known Venus co-orbitals. Source: MPC Database.
                  }
          \begin{tabular}{cccccc}
           \hline
            Object          & $\alpha$ ($^{\rm h}$:$^{\rm m}$:$^{\rm s}$) & $\delta$ (\degr:\arcmin:\arcsec) & $m$ (mag) & $\theta$(\degr) & $\phi$(\degr) \\
           \hline
            2001~CK$_{32}$  & 16:54:52.51                                 & +33:02:23.9                      & 17.2      & 83.8            & 90.0          \\
            2002~VE$_{68}$  & 01:05:00.91                                 & +26:34:22.8                      & 14.1 (R)  & 150.9           & 28.2          \\
            2012~XE$_{133}$ & 07:32:51.54                                 & +59:02:11.9                      & 18.5 (V)  & 137.0           & 40.5          \\
            2013~ND$_{15}$  & 21:21:58.739                                & -26:52:18.69                     & 20.5 (w)  & 154.3           & 22.3          \\
           \hline
          \end{tabular}
          \label{discovery}
         \end{table}
%
%
     Three out of four objects were found at elongations $>$ 135\degr. Observationally speaking, this is not unusual as NEO surveys mainly 
     focus on the regions of the sky at elongations in that range. However, and in theory, most of the objects discussed here are unlikely to 
     routinely reach perigee at such large elongations as they spend most of the time in the unobservable (daytime) sky. Morais \& 
     Morbidelli (2006) estimated that 78 per cent of Venus co-orbitals are hidden in the unobservable sky (their region III) and just three 
     per cent are found in the observable sky (their region I). In order to compute our own estimates, we performed a Monte Carlo simulation 
     (Metropolis \& Ulam 1949) in which we used the equations of the orbits of both the Earth and the object under the two-body 
     approximation (e.g. Murray \& Dermott 1999) to find the Earth-object MOID or perigee. Then we estimate the solar elongation at perigee 
     (assumed to be the most likely discovery time) and study the resulting statistics. The orbit of the Earth used in this simulation was 
     computed at Epoch JD 2456600.5 by the \textsc{horizons} system. The random values of the orbital parameters of the objects follow the 
     trends observed in Table \ref{elements}, namely 0.720~$<~a$~(au)~$<$~0.727, 0.3~$<~e~<$~0.7 and 0~$<~i~(\degr)~<$~10, with both 
     $\Omega$ and $\omega$ $\in$ (0, 360)\degr. The results of 10$^7$ test orbits are plotted in Fig. \ref{elongation} (the value of the 
     geocentric solar elongation is colour coded). Only perigees of less than 0.1 au ($\sim$91 per cent, 65 per cent are in the PHA regime) 
     are considered. The fraction of these orbits reaching perigee with elongation $<$ 90\degr (dawn to dusk sky) is 72.5 per cent; only 3.6 
     per cent of the objects reach perigee with elongations $>$ 135\degr. 
     \hfil\par
     Our results concerning the visibility of these objects are consistent with those in Morais \& Morbidelli (2006); however, our 
     interpretation is rather different. Assuming that the brightest objects are the easiest to spot and that, at elongations $>$ 135\degr, 
     the discovery efficiency of NEO surveys approaches 100 per cent (but see below), the discovery of one single bright object, 
     2002~VE$_{68}$, at elongation 150$\fdg$9 in 12 years of observations strongly suggests that at least 28 other similarly bright objects 
     are still waiting for discovery. Besides, southern declinations under -30\degr and northern declinations above +80\degr are outside the 
     coverage of most surveys. Nearly 30 per cent of the studied orbits have MOID under 0.05 au and declination under -30\degr. These facts 
     suggest that the number of asteroids with sizes above 150 m currently engaged in co-orbital motion with Venus is, at least, one order 
     of magnitude larger than usually thought. The number of smaller bodies could easily be in many thousands and they may be secondary 
     fragments, the result of tidal or rotational stresses (see e.g. Richardson, Bottke \& Love 1998). 
     \hfil\par
\begin{figure}
       \centering
        \includegraphics[width=\linewidth]{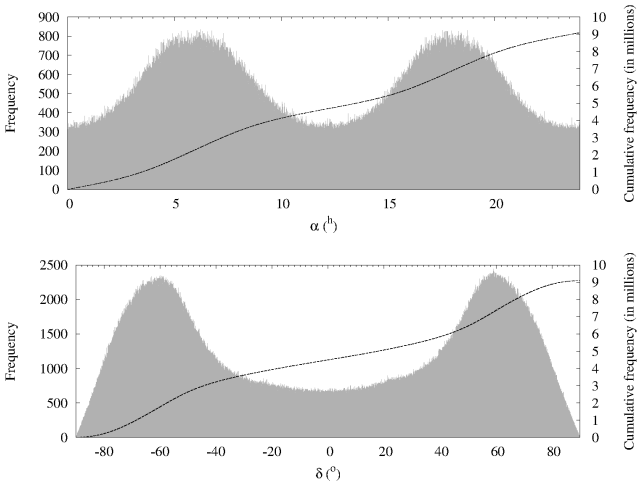}
        \caption{Frequency distribution in equatorial coordinates (right ascension, top panel, and declination, bottom panel) at perigee of
                 orbits similar to those of known Venus co-orbitals (as in Fig. \ref{elongation}). The best areas to search for these 
                 objects are located towards the ecliptic poles.
                }
        \label{radec}
     \end{figure}
\begin{figure}
       \centering
        \includegraphics[width=\linewidth]{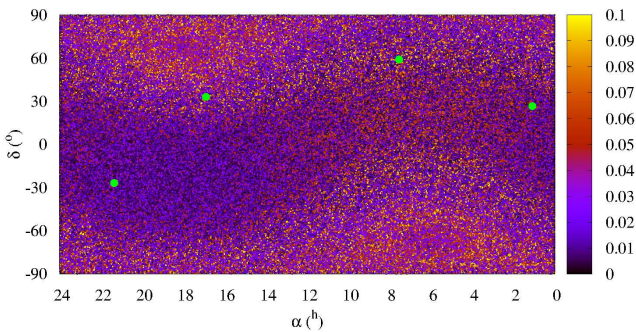}
        \caption{Distribution in equatorial coordinates of the (colour coded) perigees of objects moving in orbits similar to those of known 
                 Venus co-orbitals (as in Fig. \ref{elongation}). The green points represent the discovery coordinates of the four objects 
                 in Table \ref{discovery}.}
        \label{known}
     \end{figure}

     Fig. \ref{radec} shows the frequency distribution in equatorial coordinates at perigee for the set of orbits studied here. From this 
     and Table \ref{discovery}, it is obvious that none of the known objects were discovered within the regions of the sky where the density 
     of perigees of these objects is expected to be the highest (the ecliptic poles). Fig. \ref{known} provides the distribution in 
     equatorial coordinates of the actual perigees (colour coded). Once more it is clear that the known objects have been discovered far 
     from the optimal locations. The smallest MOIDs are found in the region within $\alpha \in (14, 22)^{\rm h}$ and $\delta \in (20, -70)$\degr.  
     Only 2013~ND$_{15}$ has been discovered close to that area of the sky. This statistical analysis can be used to implement better 
     strategies to find these objects. In any case, optimized surveys require the use of space-based telescopes. The upcoming {\it Gaia} 
     mission\footnote{http://sci.esa.int/gaia/} represents a very good opportunity to study this population and confirm/reject our current 
     views. {\it Gaia} can observe close to the Sun (down to an angular distance of 45\degr from the Sun) and should be able to observe over 77 
     per cent of these objects, if they do exist. In sharp contrast, nearly 70 per cent of the objects discussed here have elongation at 
     perigee $<$ 90\degr and, therefore, cannot be observed from the ground. The search for Trojan asteroids in the inner Solar system in 
     general, and using {\it Gaia} in particular, has been discussed by Todd, Coward \& Zadnik (2012).
     \hfil\par
%
%
%
%
%
%
%
%
%
%
%
%
%
%
     \begin{figure}
       \centering
        \includegraphics[width=\linewidth]{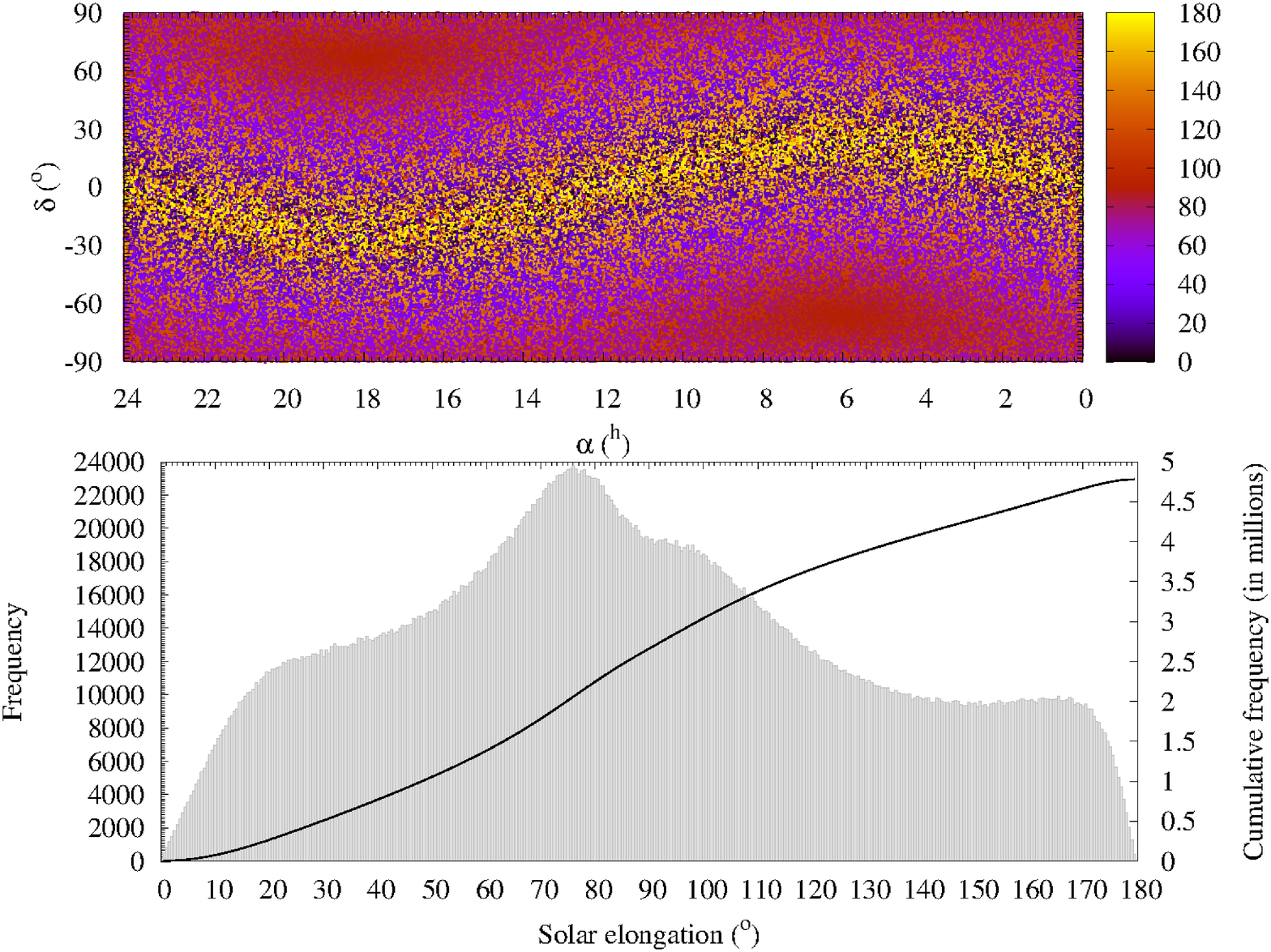}
        \caption{Similar to Fig. \ref{elongation} but for objects moving in Aten-type orbits. See the text for details. 
                 }
        \label{elongationatens}
     \end{figure}
%
%
     At this point, it may be argued that our findings are simply a corollary of the strong, but known, observational bias against detecting 
     Atens pointed out by e.g. Mainzer et al. (2012). If we perform another Monte Carlo calculation for objects moving in Aten-type orbits 
     ($a~<$~1~au, $Q~>$~0.983~au, with $Q~=~a~(1~+~e)$ being the aphelion distance, and assuming $e~<$~1.0 and $i~<$~60\degr, with both 
     $\Omega$ and $\omega$ in the range 0--360\degr) and plot the results, we obtain Fig. \ref{elongationatens}. It is true that there is indeed 
     a strong observational bias against detecting Atens. Its actual impact is however unknown because we do not know the real distribution 
     of the orbital elements of objects moving in Aten-type orbits. Our diagrams (here and in the previous figures) assume that the 
     distribution of the orbital elements is uniform but it could be the case that low-eccentricity, low-inclination orbits are less likely 
     to occur than eccentric and inclined ones due to frequent planetary close encounters. We simply do not know; a complete sample of 
     objects is needed to provide a better answer. However, if the orbital distribution is uniform or close to uniform as assumed here, the 
     predictions from our simple modelling show that the actual number of Atens should be significantly higher than the current tally. 
     Consistently, the number of objects moving in orbits similar to those of known Venus co-orbitals must be higher too but their 
     detectability is more seriously affected than that of Atens (compare frequencies in Figs \ref{elongation} and \ref{elongationatens} for 
     an equal number of test orbits, $10^7$).  
     \hfil\par
     Only 26.4 per cent of the simulated Aten-type orbits reach perigee at solar elongation $<$ 90\degr; in contrast and for hypothetical 
     Venus co-orbitals of the type discussed here, the fraction is 65.9 per cent. The difference is large enough to warrant specific 
     consideration, separated from the bulk of the Atens. Only 29.3 per cent of the Aten-type orbits studied here are moving in the PHA
     regime; for Venus co-orbitals of the type discussed here this fraction amounts to 65.1 per cent. It is obvious that the overall 
     probability of collision with our planet is much higher for the type of Venus co-orbitals discussed in this work than for the typical
     Aten population. The mean impact probabilities associated with Atens and other dynamical classes have been studied previously (see e.g. 
     Steel \& Baggaley 1985; Chyba 1993; Steel 1995; Dvorak \& Pilat-Lohinger 1999; Dvorak \& Freistetter 2001) and it is well established 
     that Atens have the shortest lifetimes against collision, closely followed by Apollos (a factor 2--5 longer). If the objects studied 
     here have lifetimes against collision with our planet even shorter than those of typical Atens, this population must be able to 
     efficiently replenish the losses via some unknown mechanism(s).
     \hfil\par
     Our analysis indicates that the existence of a relatively large population of Venus co-orbitals with sizes under 150 m is very likely.
     This small size suggests that they could be fragments of fragments. Asteroidal decay could be induced by collisional processes (e.g. 
     Ryan 2000) but also be the combined result of thermal fatigue (e.g. \v{C}apek \& Vokrouhlick\'y 2010) and tidal (e.g. T\'oth, Vere\v{s}
     \& Korno\v{s} 2011) or rotational stresses (e.g. Walsh, Richardson \& Michel 2008). These last three processes can efficiently produce 
     secondary fragments. The recent study of Mainzer et al. (2014) may point in that direction. Using NEOWISE (the asteroid-hunting portion 
     of the Wide-field Infrared Survey Explorer mission) data they have found a population of tiny NEOs characterized by increasing albedos 
     with decreasing size. This may be the result of a selection bias against finding small, dark NEOs but it may also be the specific 
     signature of a distinctive population among small objects that are the result of secondary fragmentation.

  \section{Conclusions}
     In this paper, we have identified yet another high-eccentricity Venus co-orbital and PHA, 2013~ND$_{15}$. This object is the first known
     Venus Trojan. Our calculations indicate that this small NEO moves in a transitional, highly chaotic trajectory; it has been a 
     co-orbital companion to Venus for just a few thousand years and it may become a passing object within the next few hundred years. 
     Although its future orbital evolution is not easily predictable beyond that time-scale, the object may experience recurrent co-orbital 
     episodes with Venus. However, all the calculations consistently show that 2013~ND$_{15}$ is currently in the Trojan dynamical state. 
     Due to its very eccentric orbit, its dynamical evolution is markedly different from that of the three previously known Venus 
     co-orbitals, being even less stable. On the other hand, the scarcity of detected Venus co-orbitals appears to be the result of 
     observational bias, not physical lack of objects. Our calculations suggest that the number of minor bodies with sizes above 150 m 
     currently engaged in co-orbital motion with Venus could be at least one order of magnitude larger than usually thought; the number of 
     smaller bodies could easily be in many thousands. These objects are better identified using space-based telescopes because nearly 
     70 per cent of them have solar elongation at perigee $<$ 90\degr. 

  \section*{Acknowledgements}
     The authors would like to thank S. J. Aarseth for providing one of the codes used in this research and the anonymous referee for his/her 
     constructive and useful report. This work was partially supported by the Spanish `Comunidad de Madrid' under grant CAM S2009/ESP-1496. 
     The authors also thank M. J. Fern\'andez-Figueroa, M. Rego Fern\'andez and the Department of Astrophysics of the Universidad 
     Complutense de Madrid (UCM) for providing computing facilities. Most of the calculations and part of the data analysis were completed 
     on the `Servidor Central de C\'alculo' of the UCM. In preparation of this paper, the authors made use of the NASA Astrophysics Data 
     System, the ASTRO-PH e-print server, the MPC data server and the NEODys service.

\end{document}